# Yielding and hardening of flexible fiber packings during triaxial compression

Yu Guo[1*], Yanjie Li[2], Qingzhao Liu[2], Hanhui Jin[1], Dandan Xu[3], Carl Wassgren[4], Jennifer S. Curtis[5]

[1]Department of Engineering Mechanics, Zhejiang University, Hangzhou, 310027, China
[2]School of Technology, Beijing Forestry University, Beijing 100083, China
[3]College of Mechanical Engineering, Zhejiang University of Technology, Hangzhou, 310023, China
[4]School of Mechanical Engineering, Purdue University, West Lafayette, IN 47907, USA
[5]Department of Chemical Engineering, University of California Davis, Davis, CA 95616, USA
[*]Corresponding author: yguo@zju.edu.cn

**Abstract**

This paper examines the mechanical response of flexible fiber packings subject to triaxial compression. Short fibers yield in a manner similar to typical granular materials in which the deviatoric stress remains nearly constant with increasing strain after reaching a peak value. Interestingly, long fibers exhibit a hardening behavior, where the stress increases rapidly with increasing strain at large strains and the packing density continuously increases. Phase diagrams for classifying the bulk mechanical response as yielding, hardening, or a transition regime are generated as a function of the fiber aspect ratio, fiber-fiber friction coefficient, and confining pressure. Large fiber aspect ratio, large fiber-fiber friction coefficient, and large confining pressure promote hardening behavior. The hardening packings can support much larger loads than the yielding packings contributing to the stability and consolidation of the granular structure, but larger internal axial forces occur within fibers.

**Main text**

Triaxial compression, in which the sample is compressed in one direction and confined with a constant pressure in the directions perpendicular to the compression direction, has been widely used to measure shear yielding properties of various materials. In triaxial compression of common granular materials (sands, for example), the deviatoric stress initially increases to a peak value and then drops or remains nearly constant, with fluctuations, as the strain increases [1-5]. After reaching the peak stress (usually called a yield stress), a granular material yields and undergoes plastic deformation. Recent x-ray experiments in a sheared spherical particle bed show the microstructural change in the plastic yielding process [6]. It was found that highly distorted tetrahedra are the structural defects for microscopic plasticity, and plastic deformation occurs through flip events of these unstable coplanar tetrahedra in the microstructure [6]. This continuous microstructural change contributes to the yielding behavior. The magnitude of the yield stress depends on the packing density, particle size, particle shape, and the inter-particle contacts. A larger packing density and larger particle size lead to a larger yield stress [7]. The effect of particle size distribution on yield stress is complex and depends on the shape of the distribution curve [8, 9] since the interaction between different particle species affects the overall stress. Particles of more angular shape (polyhedral particles, for example) have an increased chance of interlocking, causing a more anisotropic contact force network and a larger yield stress [2, 10, 11]. More significant stress fluctuations are observed for flatter particles compared to rounder and more prolate particles due to the occurrence of avalanche-like slip events [5]. Large friction coefficients, large rolling resistance, and strong cohesive interactions between particles constrain the relative particle movement and therefore increase the yield stress [12-14].

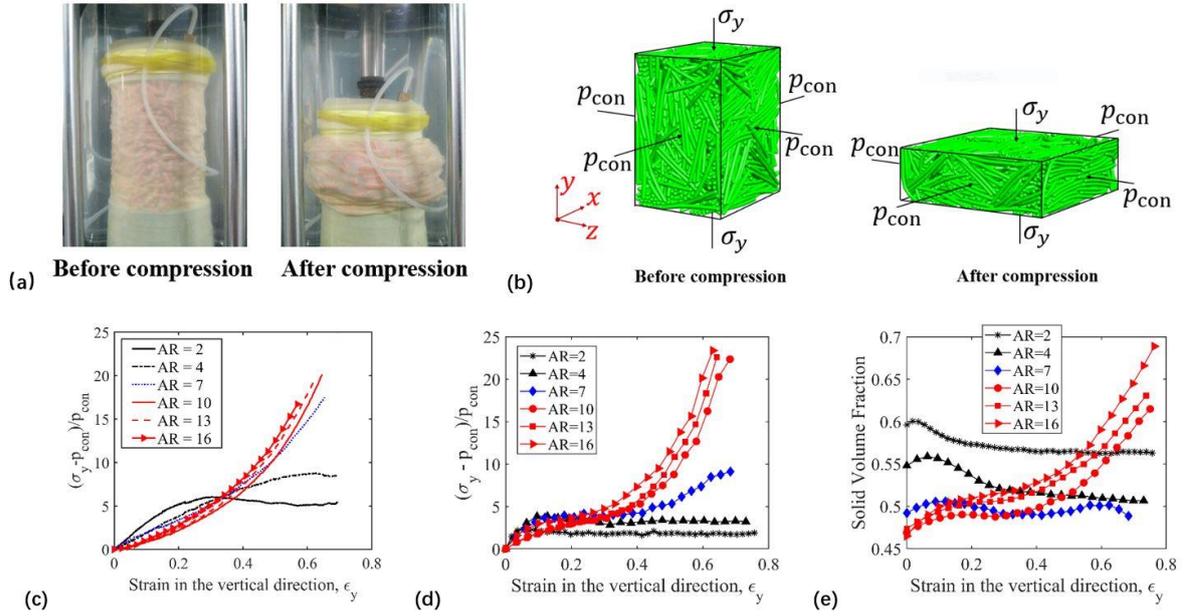

**Figure 1.** (a) Experimental set-up and (b) numerical models for triaxial compression of an assembly of flexible fibers: the states before and after compression. (c) Experimental results and (d) simulation results of the normalized deviatoric stresses, $(\sigma_y - p_{con})/p_{con}$, varying with the strain in the vertical direction, $\epsilon_y$. (e) Solid volume fraction as a function of strain $\epsilon_y$ from the simulations. The confining pressure, $p_{con}$, is set to 10 kPa and the fiber-fiber friction coefficient is 1.4 in the experiments and simulations. Various fiber aspect ratios, *AR*, are considered. In the diagrams (c)-(e), each curve is obtained based on the average results of at least three independent tests with different initial configurations of fiber packings.

In the biaxial compression of elongated rod-like particles [15], the yield stress increases with increasing particle aspect ratio, defined as the ratio of the particle's length to diameter. As the particles become more elongated, the proportion of sliding contacts increases significantly and the local nematic ordering occurs with the particle major axes oriented perpendicular to the major principal stress direction. In these previous studies, the rod-like particles were rigid and the maximum particle aspect ratio was not greater than four [15]. In this study, we performed triaxial compression experiments and three-dimensional simulations with flexible fibers, which can undergo stretching, bending, and twisting deformations and have much larger aspect ratios (up to 25). We observe that when the fiber aspect ratio is sufficiently large, the bulk material does not yield and the deviatoric stress increases with increasing strain in a power law relationship. This hardening behavior during triaxial compression is caused by the interlocking of elongated fibers, which forms a stable, irregular knitted structure. Thus, due to hardening, a packing of elongated fibers can potentially support a much greater load than a packing of typical granular material subject to the same confining pressure. The major aim of this work is to explore the conditions under which yielding or hardening occurs for various fiber packings.

Experiments, as shown in Figure 1a, are conducted on a TSZ-3A strain-controlled triaxial test apparatus (Nanjing Soil Instrument). A sample of cylindrical shape is created by filling fibers into a latex film of 0.3 mm thickness. The sample, which has a diameter of 61.8 mm and a height of 50-90 mm, is fully submerged into a pool of water, which provides a constant hydraulic pressure, $p_{\text{con}}$. A piston moves downwards at a constant speed of 2.5 mm/min to compress the sample, causing the sample to expand in the radial direction. A thin pipe connected to a hole on top of the sample is used to release the air inside the sample, thus avoiding a build-up of air pressure. The normal stress exerted on the bottom of the piston by the sample, $\sigma_y$, is measured during the compression process. The fibers used in the experiments are silicon rubber flexible cords of diameter $d$ = 2.4 mm, density $\rho$ = 1261 kg/m³, cord-cord friction coefficient $\mu$ = 1.4, and Young's modulus $E$ = 6.35 MPa. Various lengths (i.e., various aspect ratios) of rubber cords are employed in the tests.

Numerical simulations of flexible fiber triaxial compression are also performed using the Discrete Element Method (DEM), because some information, such as the fiber-fiber contact forces and internal forces within individual fibers, can be easily obtained from the numerical simulations. A fiber is formed by elastically bonding sphero-cylinders in a straight line. In principal, the discretized numerical fiber model is closer to a continuum fiber as the number of elemental sphero-cylinders increases. It is found that the stress-strain results of triaxial compressions are converged with increasing number of elements in a fiber. The number of elements is 10 for fibers of aspect ratio 25. Increasing the number of sphero-cylinder elements produces nearly the same stress-strain results. The fiber exhibits deformation as the elastic bonds undergo stretching/compressing, shearing, bending, and twisting deformations. The bond force and moment exerted on the sphero-cylinder elements are a function of the corresponding bond deformation. The mathematical details and validation of the flexible fiber model are provided in previous work [16]. The contact calculations between fibers are based on sphero-cylinders. The contact force model for sphero-cylinders proposed by Kidokoro et al. [17] has been employed for the normal contact and the Mindlin model [18] has been used for the tangential contact force. The computational set-up is shown in Figure 1b. Sample preparation proceeds as follows. A number of fibers are randomly generated in a cube confined by six planar walls. A constant confining pressure, $p_{\text{con}}$, is applied to the top plane and four lateral planes, which can only move in the corresponding pressure directions. The bottom plane is fixed. After the system comes to rest, the top plane moves downwards at a slow speed to compress the sample. The confining pressure on the four lateral planes remains constant in the compression process. The normal stress exerted on the top plane, $\sigma_y$, is obtained from the simulations. The same numerical setup has been employed in previous studies using rigid polyhedral particles [3]. The material properties of the silicon rubber cords used in the experiments are employed as the input parameters for the DEM simulations. In the simulations, the total mass is the same for each fiber bed with a specified fiber aspect ratio. The total number of fibers used in a triaxial compression simulation is sufficient and more fibers produce the same yielding or hardening behavior. The total number of flexible fibers of aspect ratio 25 in the triaxial compression simulation is 550.

Experimental and simulation results of the normalized deviatoric stresses, $(\sigma_y - p_{\text{con}})/p_{\text{con}}$, varying with the strain in the vertical direction, $\epsilon_y$, are shown in Figure 1c and 1d, respectively. The confining pressure, $p_{\text{con}}$, is set to 10 kPa. The shorter fibers with aspect ratios *AR* = 2 and 4 show similar yielding behavior as typical granular materials: the stress initially increases and then remains constant with increasing strain. In contrast, fibers with larger aspect ratios show no yielding. Instead, the stress and strain follow a power law relationship,

$$\frac{\sigma_y - p_{\text{con}}}{p_{\text{con}}} \propto \epsilon_y^c \tag{1}$$

For the fibers with AR =7, the power law constant $c$ in Eq. (1) is close to one, indicating a linear relationship between stress and strain. For the fibers with AR $\geq$ 10, the power law constant $c$ is greater than one, indicating that the sample stiffness increases with increasing strain. Thus, the mechanical responses of a packing of fibers can be classified into three regimes: yielding with the power law constant $c$ equal to zero, transition with $c$ between zero and one, and hardening with $c$ greater than one. As the fiber aspect ratio increases, the mechanical response of a packing of fibers subject to triaxial compression varies from yielding, through transitional behavior, to hardening. As shown in Figure 1e, the solid volume fraction during the compression process exhibits different trends for the different mechanical responses: i) For yielding, the fiber packing generally dilates and the solid volume fraction decreases and converges to a lower limit; ii) For the transitional response, the solid volume fraction shows little change; iii) For hardening, the packing of fibers continuously consolidates and the solid volume fraction increases as the compression proceeds.

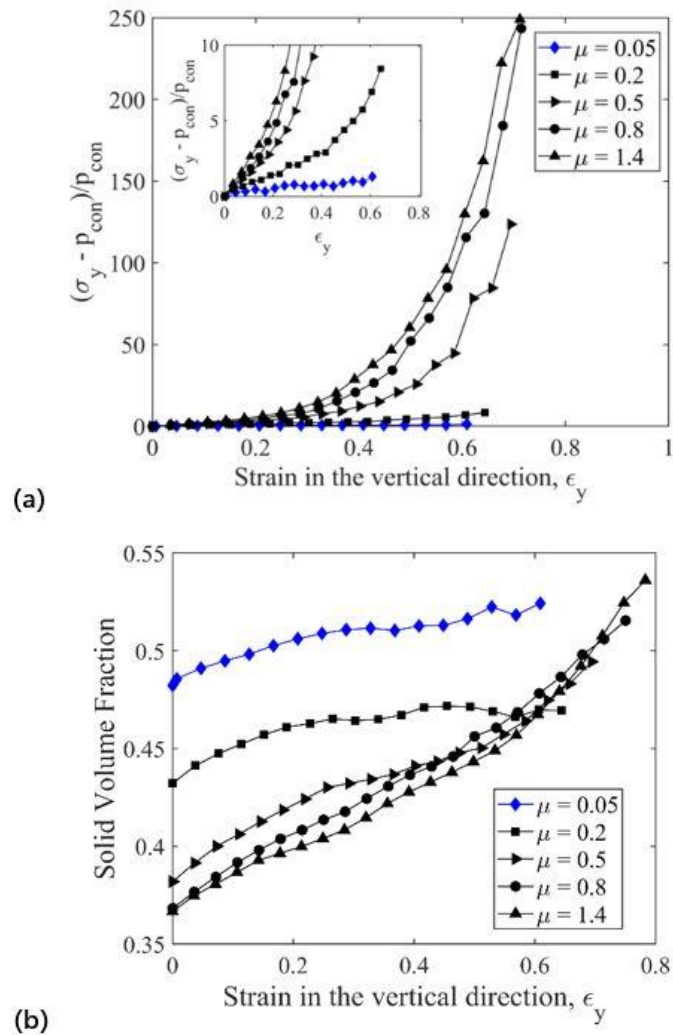

**Figure 2**. Effect of fiber-fiber friction coefficient, $\mu$, on the (a) normalized deviatoric stresses and (b) solid volume fraction. The confining pressure, $p_{con}$, is set to 1 kPa and the fiber aspect ratio, *AR*, is 25. The results are obtained from the DEM simulations. Each curve is obtained based on the average results of three independent tests with different initial configurations of fiber packings.

The fiber-fiber friction coefficient also has a critical impact on the mechanical behavior of the fiber packing. As shown in Figure 2a, material hardening with $c > 1$ is observed for larger fiber-fiber friction coefficients, e.g., $\mu \geq 0.2$, and transitional behavior with $c = 1$ is observed for small fiber-fiber friction coefficients, e.g., $\mu = 0.05$. Thus, a large inter-fiber friction coefficient promotes material hardening. For the initial packing before compaction, a larger solid volume fraction is obtained for the less frictional fibers due to the fact that smoother fibers more easily slide past their neighbors to fill voids. During compression, the solid volume fraction increases slightly for the fibers with smaller friction coefficients ($\mu = 0.05$ and $0.2$) and increases significantly for fibers with larger friction coefficients, which show macroscopic hardening behavior.

It is evident that the mechanical response of a packing of fibers depends on the fiber aspect ratio and fiber-fiber friction coefficient. Therefore, phase diagrams of yielding, transitional, and hardening regimes based on the fiber aspect ratio, *AR*, and fiber-fiber friction coefficient, *μ*, for two different confining pressures, $p_{con}$, are plotted in Figure 3. It can be seen that material yielding occurs for fibers with small aspect ratios while hardening occurs for the fibers with large aspect ratios. A larger fiber-fiber friction coefficient acts to reduce the size of the yielding and transitional regimes. By comparing Figures 3a and 3b, the confining pressure, $p_{con}$, is observed to affect the positions of the boundaries between different regimes. The boundary between the yielding and transitional regimes and the boundary between the transitional and hardening regimes moves toward smaller aspect ratios as the confining pressure increases. Figure 3b shows that the simulation predictions of the regimes for the fibers with the friction coefficient *μ* = 1.4 are in good agreement with the experimental results.

The hardening of a fiber packing occurs because elongated, frictional fibers become entangled and interlocked to form strong structures that are capable of supporting large loads. When a fiber packing is compressed by external loads, individual fibers are stretched or compressed along the fiber axes, inducing significant internal fiber forces. From the DEM simulations, the bond forces holding the elements together to form a fiber can be obtained. The bond forces represent the internal fiber forces. Probability density functions (PDF), $P$, of normalized bond axial forces, $f_b/(E_b A)$, with various fiber aspect ratios and a fixed fiber-fiber friction coefficient of $\mu$ = 1.4 are shown in Figure 4a. The parameter $E_b$ is the Young's modulus of the bonds and $A$ is the cross-sectional area of the fiber. The positive and negative signs reflect bond tensile and compressive forces, respectively. The distributions become wider and a larger portion of large bond forces are obtained for larger aspect ratio fibers. Thus, larger bond forces exist in the larger aspect ratio fibers exhibiting hardening behavior. Probability density functions with various friction coefficients and a fixed fiber aspect ratio of *AR* = 13 are shown in Figure 4b. By increasing the fiber-fiber friction coefficient, the bond forces increase and the mechanical response changes from transitional to hardening (see Figure 3a).

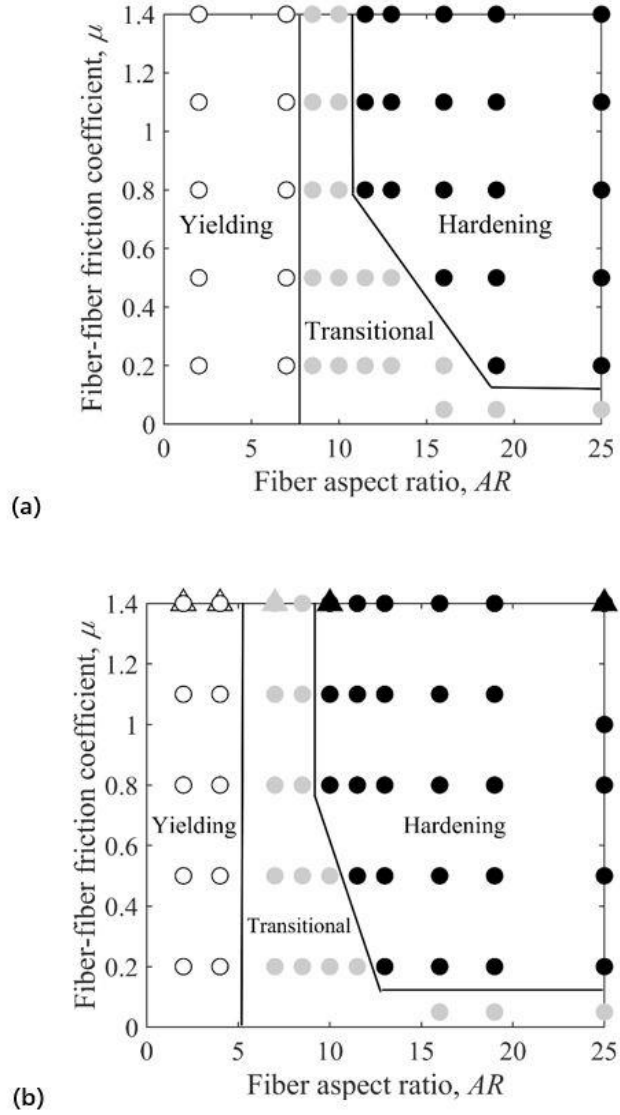

**Figure 3.** Phase diagrams showing yielding, transitional, and hardening regimes based on the fiber aspect ratio, $AR$, and fiber-fiber friction coefficient, $\mu$, for the confining pressures, $p_{con}$, of (a) 1 kPa and (b) 10 kPa. The circle symbols are the simulation results and the triangle symbols are the experimental results.

For the bond tensile forces larger than the average value, the probability density function, $P$, follows an exponential relationship with the normalized bond force,

$$P = e^{-\beta_b f_b/(E_b A)} \tag{2}$$

where $\beta_b$ represents the magnitude of the slope of the distribution line above the average value of $f_\mathrm{b}/(E_\mathrm{b}A)$ in Figures 4a and 4b. A smaller value of $\beta_\mathrm{b}$ indicates a greater portion of larger bond tensile forces. The variation of $\beta_\mathrm{b}$ with the fiber aspect ratio and fiber-fiber friction coefficient is shown in Figure 4c. The values of $\beta_\mathrm{b}$ are determined by fitting each set of data above the average value of normalized tensile forces $f_\mathrm{b}/(E_\mathrm{b}A)$ in Figures 4a and 4b. The plot demonstrates that the distribution parameter $\beta_\mathrm{b}$ decreases and therefore the tensile fiber forces increase with increasing fiber aspect ratio and friction coefficient. It is noted that the distributions of fiber-fiber contact forces (not shown here) are similar to those of the tensile fiber forces (Figures 4a and 4b). Also, fiber aspect ratio and friction coefficient have similar effects on the fiber-fiber contact forces and internal fiber forces. Thus, fiber packings exhibiting hardening behavior have large internal axial forces within fibers and also large fiber-fiber contact forces.

We emphasize that the macroscopic responses of flexible fiber packings in triaxial compression can be classified into three regimes: yielding, transitional, and hardening. Both the deviatoric stress and solid volume fraction increase with increasing strain for the hardening fiber packings. The phase diagrams indicate that the different regimes depend on the fiber aspect ratio, fiber-fiber friction coefficient, and confining pressure. Larger aspect ratio, friction coefficient, and confining pressure promote hardening behavior. It is speculated that hardening is caused by the entanglement and interlocking of elongated, frictional fibers. The DEM simulations results show that much larger internal axial forces (including tensile and compressive) within the individual fibers occur for the hardening packings than for the yielding packings, indicating that the improved bearing capacity of the hardening packings is achieved by inducing larger internal fiber forces. Future work will examine the role of fiber stiffness, which impacts individual fiber deformation and will likely affect the regimes and transitions shown in Figure 3.

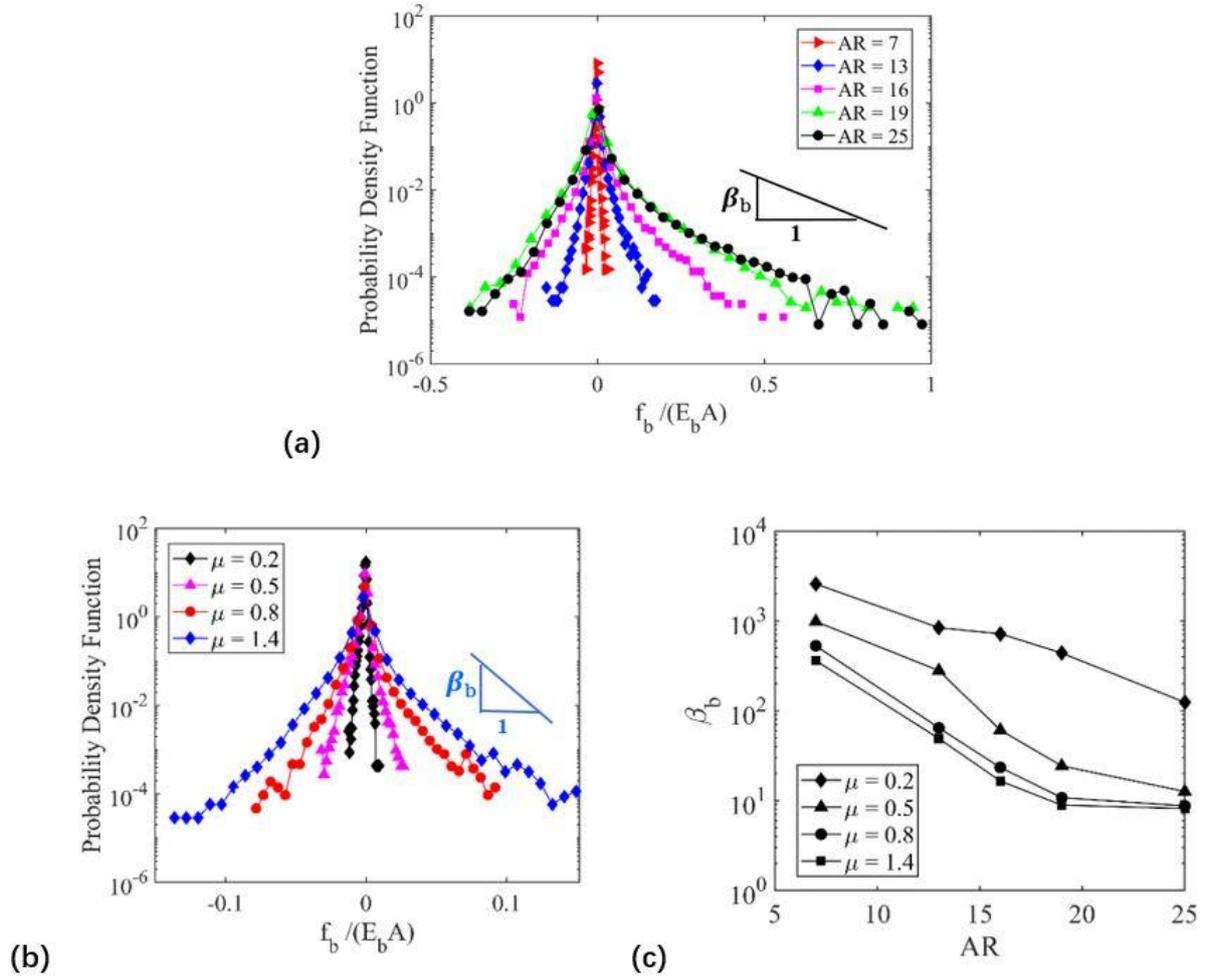

**Figure 4.** Probability density functions of normalized bond normal forces with (a) various fiber aspect ratios, *AR,* and a fiber-fiber friction coefficient of $\mu$ = 1.4 and (b) various friction coefficients $\mu$ and a fiber aspect ratio of *AR* = 13. The dependence of the exponential distribution parameter, $\beta_b$, on *AR* and $\mu$ is shown in (c). The confining pressure, $p_{\text{con}}$, is set to 1 kPa. The results are obtained from the DEM simulations.

**Acknowledgement**

The National Science Foundation of China (Grant NO. 11872333) and the Zhejiang Provincial Natural Science Foundation of China (Grant NO. LR19A020001) are acknowledged for the financial supports.